\documentclass[twocolumn,groupedaddress]{revtex4}
\usepackage{graphicx}

\begin{document}

% Use the \preprint command to place your local institutional report
% number in the upper righthand corner of the title page in preprint mode.
% Multiple \preprint commands are allowed.
% Use the 'preprintnumbers' class option to override journal defaults
% to display numbers if necessary
%\preprint{}

%Note \\ will produce line breaks
%Title of paper
\title{Separation of Scales in the More Effective Field Theory \\ and Moszkowski-Scott Methods}

% repeat the \author .. \affiliation  etc. as needed
% \email, \thanks, \homepage, \altaffiliation all apply to the current
% author. Explanatory text should go in the []'s, actual e-mail
% address or url should go in the {}'s for \email and \homepage.
% Please use the appropriate macro foreach each type of information

% \affiliation command applies to all authors since the last
% \affiliation command. The \affiliation command should follow the
% other information
% \affiliation can be followed by \email, \homepage, \thanks as well.
\author{J.W. Holt}
%\email{jeholt@grad.physics.sunysb.edu}
\author{G.E. Brown}
%\homepage[]{Your web page}
%\thanks{}
%\altaffiliation{}
\affiliation{Department of Physics, SUNY, Stony Brook, New York 11794, USA}

%Collaboration name if desired (requires use of superscriptaddress
%option in \documentclass). \noaffiliation is required (may also be
%used with the \author command).
%\collaboration can be followed by \email, \homepage, \thanks as well.
%\collaboration{}
%\noaffiliation

\date{\today}

\begin{abstract}
We compare the momentum-space decimation procedure used to construct the low momentum nucleon-nucleon interaction $V_{\rm low-k}$ with the configuration-space separation method of Moszkowski and Scott. Each procedure defines a separation of scales in the nuclear many-body problem, and the extent to which these two scales coincide is studied. By studying the effects of the separation method on the relative $S$-state Kallio-Kolltveit potential, it is found that close agreement with $V_{\rm low-k}$ is obtained as the configuration-space cutoff is lowered to \mbox{$\sim 1.0$ fm}.
\end{abstract}

% insert suggested PACS numbers in braces on next line
\pacs{}
% insert suggested keywords - APS authors don't need to do this
%\keywords{}

%\maketitle must follow title, authors, abstract, \pacs, and \keywords
\maketitle

%_________________________________________________________________________________

\section{Introduction}
The {\it more effective} effective field theory (MEEFT) approach to constructing a low momentum effective nucleon-nucleon (NN) interaction has proven to be a highly successful procedure. By introducing a momentum space cutoff $\Lambda$, it has been shown \cite{Bog1,Bog2} that all high precision NN potentials that reproduce the experimental phase shift data up to \mbox{$E_{\rm lab} \simeq 350$ MeV} flow to a nearly unique interaction $V_{\rm low-k}$ as the cutoff is lowered to \mbox{$\Lambda \simeq 2$ $\rm fm^{-1}$}. Removing the large momentum modes of an interaction corresponds to removing the short distance details, but the exact extent to which the MEEFT procedure removes these short distance details is not well understood. In particular, it would be useful to understand whether the momentum space cutoff of \mbox{$\sim 2$ $\rm fm^{-1}$} corresponds to an approximate cutoff in position space. The purpose of the present paper is to investigate this question.

The MEEFT, which is reviewed in \cite{Brown1}, is renormalization group (RG) ``friendly''. It is {\it more effective} in the sense that the cutoff $\Lambda$ is chosen so as to include all experimental data that have been converted into precision NN potentials. Since the maximum momentum in the data corresponds to a cms momentum \mbox{$\Lambda \simeq 2$ $\rm fm^{-1}$}, it makes no sense to increase $\Lambda$, which would then include contested inner parts of the potentials, or to decrease $\Lambda$ (decimate), which would mean cutting out some of the experimental data. With $\Lambda$ chosen as it is, all well measured and well analyzed data are included in the (unique) $V_{\rm low-k}$.

Removing the large momentum or short distance details of an interaction in order to construct an effective interaction is not, however, a new tool in nuclear physics. The separation method of Moszkowski and Scott \cite{Mos} provided 40 years ago a means by which the nuclear interaction is uniquely divided into a short distance potential and a long distance potential. This separation is made in such a way that the short distance potential gives no phase shift for free particle scattering; the long distance potential is then used as a first approximation to the effective interaction in nuclear matter. Both the separation method and MEEFT establish a separation of scales in the nuclear interaction, the former in configuration space and the latter in momentum space. The extent to which these two scales coincide will be the primary investigation of this paper.

We have chosen to compare the MEEFT procedure and the separation method by way of the Kallio-Kolltveit (KK) potential \cite{Kal1,Kal2}, a relative $S$-state potential that has been chosen primarily for its simplicity. By comparing the end products of the MEEFT procedure and the MS separation method, it is hoped that some interesting, semi-quantitative connections can be established between the two methods. In particular, we will draw conclusions regarding the extent of locality in configuration space of $V_{\rm low-k}$.

%_______________________________________________________________________________
\section{MEEFT and the NN Interaction}
Because the nuclear force cannot at present be derived from the underlying theory of QCD, a number of phenomenological meson-exchange models have been developed to describe the NN interaction. At large distances all of these potentials have the one-pion-exchange character, but at intermediate and short distances they differ significantly. Despite these differences, all of the high precision potentials correctly reproduce the experimentally observed deuteron binding energy and low energy nucleon phase shift data. To remove the model dependence in the NN interaction, the renormalization group is used in MEEFT to construct a unique low momentum effective interaction $V_{\rm low-k}$ according to the procedure described below.

A principle requirement of any RG procedure is that low energy observables--in this case the deuteron binding energy and low energy $T$-matrix--be preserved under the RG transformation. So, beginning with the full-space half-on-shell $T$-matrix
\begin{eqnarray}
\nonumber &&T(k',k,k^2) = V_{NN}(k',k) \\
 & & {} + \frac{2}{\pi}{\cal P} \int _0 ^{\infty} \frac{V_{NN}(k',q)T(q,k,k^2)}{k^2-q^2} q^2 dq, \\ \nonumber
\end{eqnarray}
we define a low-momentum half-on-shell T-matrix by
\begin{eqnarray}
\nonumber &&T_{\rm low-k }(p',p,p^2) = V_{\rm low-k }(p',p) \, + \\
& & {} + \frac{2}{\pi}{\cal P} \int _0 ^{\Lambda} \frac{V_{\rm low-k }(p',q) T_{\rm low-k} (q,p,p^2)}{p^2-q^2} q^2 dq, \\ \nonumber
\end{eqnarray}
where $\cal P$ denotes the principal value and the cutoff $\Lambda$ will be taken to be 2.0 $\rm fm^{-1}$. These two $T$-matrices are required to be identical for momenta $p < \Lambda$, and it can be shown \cite{Bog3} that a $V_{\rm low-k}$ defined by 
\begin{eqnarray}
\nonumber && V_{\rm low-k} = \hat{Q} - \hat{Q'} \int \hat{Q} \\
& & {} + \hat{Q'} \int \hat{Q} \int \hat{Q} {} - \hat{Q'} \int \hat{Q} \int \hat{Q} \int \hat{Q} + \cdots \\ \nonumber
\end{eqnarray}
will satisfy this requirement. In the above equation, $\hat{Q}$ is an irreducible vertex function and $\hat{Q'}$ is obtained by removing from $\hat{Q}$ all terms first order in the interaction $V_{NN}$. There are several schemes \cite{And,Kren} available for accurately computing $V_{\rm low-k}$, and each scheme preserves the deuteron binding energy. Under this RG procedure, all of the high precision $V_{NN}$ flow, as \mbox{$\Lambda \rightarrow 2.0$ $\rm fm^{-1}$}, to a nearly unique interaction $V_{\rm low-k}$, whose relative $S$-states will be a subject of analysis later in the paper.

%________________________________________________________________________________
\section{The Separation Method}
Moszkowski and Scott introduced the separation method to simplify and illuminate Brueckner's approach to the nuclear many-body problem. In Brueckner theory one introduces a reaction matrix, $G$, whose diagonal elements are defined by
\begin{eqnarray}
\nonumber \lefteqn{\left \langle ij |G| ij \right \rangle = \left \langle ij |V| ij \right \rangle} \\
& & {} + \sum_{m,n>k_f} \frac{\left \langle ij |V| mn \right \rangle \left \langle mn |G| ij \right \rangle}{\epsilon_i + \epsilon_j - \epsilon_m - \epsilon_n}, \\ \nonumber
\end{eqnarray}
or in operator notation
\begin{equation}
G = V + V\frac{Q}{e}G,
\label{react}
\end{equation}
where $Q$ is a Pauli operator that prevents scattering into occupied states and $e$ is the energy denominator. Equation (\ref{react}) can be expanded as an infinite series, and a systematic discussion of higher-order corrections has been discussed in \cite{Bethe}. The purpose of the reaction matrix is to deal with the difficulties associated with the nuclear hard core. This hard core makes any kind of treatment by perturbation theory impossible, but by introducing the reaction matrix it is possible to calculate the total energy of the many-body system according to 
\begin{equation}
E = \sum_{m} T_m + \frac{1}{2} \sum_{m, n} \left \langle mn |G| mn - nm \right \rangle,
\end{equation}
where all of the sums are performed over just the occupied states. Calculations \cite{Brueck} performed within the framework of Brueckner theory show that the two-particle relative wavefunction in nuclear matter is essentially equal to the unperturbed wavefunction beyond \mbox{$\sim 1$ fm}. Because of the Pauli exclusion principle, it must ``heal'' to the unperturbed wavefunction at a distance of \mbox{$\sim k_F^{-1} \simeq (2m_\pi)^{-1}$}. The separation method of Moszkowski and Scott provides an illuminating derivation of this fact and has been used in \cite{Mos} to derive an alternate expansion for the reaction matrix that converges more rapidly than (\ref{react}).

\begin{figure}[htbp]
\centerline{\includegraphics[width=5cm, height=4cm]{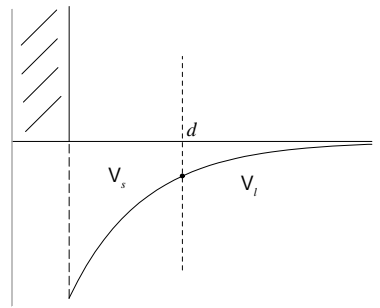}}
\caption{Separation of the NN potential into short and long distance parts.}
\label{cutoff}
\end{figure}

The essential idea of the separation method is to ``cancel'' the problematic hard core with part of the short-distance attractive well and use the remaining long-distance part as the effective interaction in nuclear matter. A repulsive interaction--even an infinite hard core--produces a finite negative phase shift, whereas an attractive potential produces a positive phase shift. For incident energies that are not too large, the nuclear potential leads to an overall positive phase shift. In these cases it is possible to combine the repulsive core with the attractive well up to a distance $d$ such that the combination of the two, called $V_s$, will give zero phase shift for free particle scattering. The remaining part of the potential, called $V_l$, will then produce the same free particle phase shifts as the original, but without the presence of the hard core. Figure \ref{cutoff} shows the separation \mbox{$V_{NN} = V_s + V_l$} for a general NN potential. 
Under this definition of the separation distance, at \mbox{$r = d$}
\begin{equation}
\frac{1}{r\Psi}\frac{d(r\Psi)}{dr} = \frac{1}{r\phi}\frac{d(r\phi)}{dr},
\end{equation}
where $\Psi$ is the free-space relative wavefunction and $\phi$ is the unperturbed wavefunction. Enforcing this criterion is how the separation distance is calculated in practice.

The NN potential depends on the relative angular momentum of the two nucleons, so the separation distance will in general depend on the relative angular momentum state. Furthermore, free particle phase shifts are momentum dependent, and therefore the separation distance will also be a function of the relative momentum. In practice, one can work with a separation distance that is momentum dependent or else fix the separation distance and include corrections. Indeed, in the Kuo-Brown interaction \cite{Kuo} the separation method was used to construct a momentum-independent separation distance for the individual $S$-states, but the Reference Spectrum method \cite{Bethe} was convenient for states of other angular momenta.

We now summarize the main results reached by Moszkowski and Scott. By separating the potential into $V_s$ and $V_l$ in the manner described above, it can be argued in a qualitative manner that the in-medium relative wavefunction is approximately equal to the free-space relative wavefunction below the separation distance $d$. In other words, effects due to nuclear matter are relatively weak below $d$. Furthermore, the relative wavefunction beyond $d$ can be shown to be approximately equal to the unperturbed wavefunction--in agreement with the results of Brueckner and Gammel \cite{Brueck}. Finally, under these two approximations for the relative wavefunction in nuclear matter, to first order the diagonal elements of the $G$-matrix be simply equal to the diagonal elements of $V_l$. Mathematically rigorous arguments performed within Brueckner theory show that the $G$-matrix is more accurately given by
\begin{eqnarray}
\nonumber &&G = V_l + G_s + G_s\frac{Q-1}{e}G_s + G_s\left (\frac{1}{e}-\frac{1}{e_0}\right )G_s \\
& & {} + V_l\frac{Q}{e}G_s + G_s\frac{Q}{e}V_l + V_l\frac{Q}{e}V_l + \cdots, \\ \nonumber
\end{eqnarray}
where $G_s$ is the $G$-matrix for $V_s$ alone and $e_0$ is the free particle propagator. Through direct calculation Moszkowski and Scott showed that the terms in this series converge more rapidly than those in (\ref{react}). For our purposes, though, we will use the approximation \mbox{$G \simeq V_l$}.

%_____________________________________________________________________________________
\section{Kallio-Kolltveit Potential}
Restricting their attention to relative $S$-states, Kallio and Kolltveit modeled the free space NN interaction with a potential of the form
\begin{equation}
V(r) = \frac{3+\sigma_1 \cdot \sigma_2}{4}V_t(r) + \frac{1-\sigma_1 \cdot \sigma_2}{4}V_s(r),
\end{equation}
where 
\begin{equation}
V_i(r) = \left \{ \begin{array}{cc}
	\infty & \mbox{for $r \leq 0.4$ fm} \\
	-A_ie^{-\alpha_i(r-0.4)} & \mbox{for $r > 0.4$ fm} \\
	\end{array} \right.  \mbox{ for $i = s,t$.}
\end{equation}
The four parameters were determined by fitting the scattering length and effective range:
\begin{equation}
\begin{array}{cc}
	A_s = 330.8 \mbox{ MeV}, & \alpha_s = 2.4021 \mbox{ $\rm fm^{-1}$} \\
	A_t = 475.0 \mbox{ MeV}, & \alpha_t = 2.5214 \mbox{ $\rm fm^{-1}$} \\
\end{array}
\end{equation}
It is important to include the infinite hard core, known to schematize the vector meson exchange, in the effective potential. To compensate for this the potentials must be very attractive, as shown by $A_s$ and $A_t$, so that the separation distance will change only slowly with incident energy. Applying the separation method to this potential yields singlet and triplet separation distances of \mbox{$d_s=1.025$ fm} and \mbox{$d_t=0.925$ fm}, which have been shown to vary slowly with relative momentum \cite{Kal1}.

%___________________________________________________________________________________

\section{Results and Discussion}
\begin{figure}[htbp]
\centerline{\includegraphics[width=6.5cm, height=5.5cm]{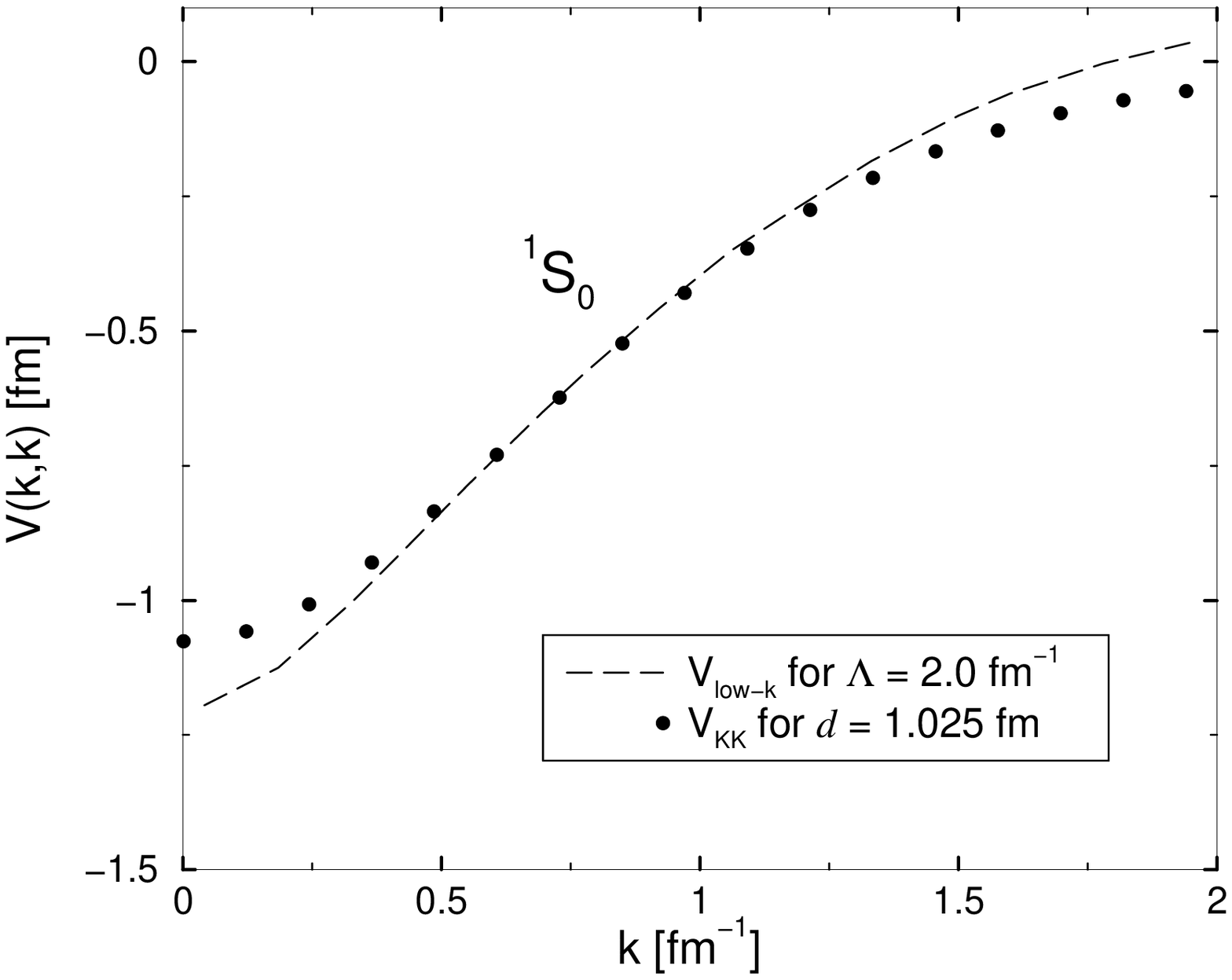}}
\caption{The $^1S_0$ diagonal matrix elements of $V_{\rm low-k}$ and the Kallio-Kolltveit potential for a configuration space cutoff of \mbox{1.025 fm}.}
\label{1s0}
\end{figure}

\begin{figure}[htbp]
\centerline{\includegraphics[width=6.5cm, height=5.5cm]{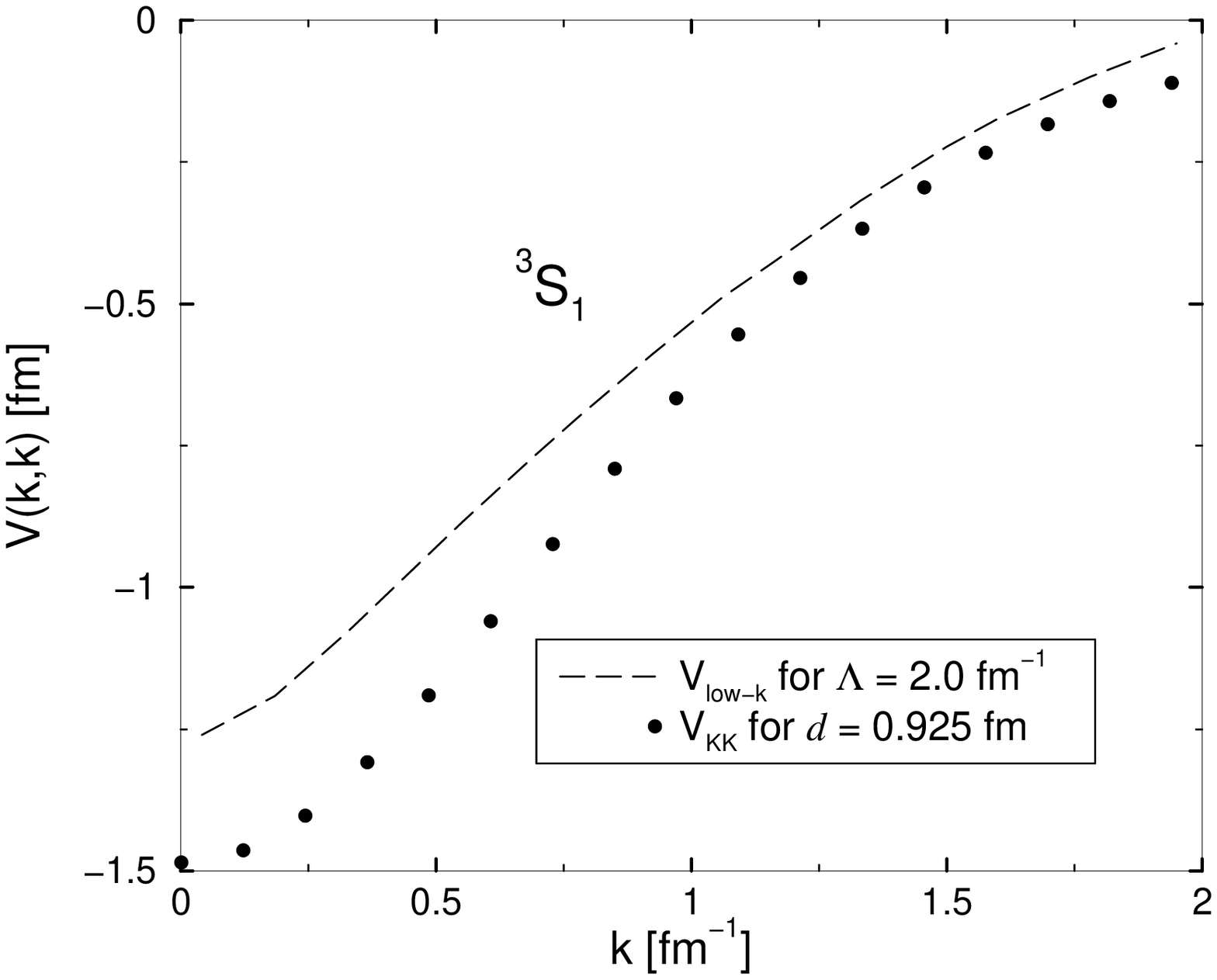}}
\caption{The $^3S_1$ diagonal matrix elements of $V_{\rm low-k}$ and the Kallio-Kolltveit potential for a configuration space cutoff of \mbox{0.925 fm}.}
\label{3s1a}
\end{figure}

In order to directly compare the KK potential with $V_{\rm low-k}$, it is convenient to transform the KK potential into $k$-space according to
\begin{equation}
V^l(k,k^\prime) = \frac{2}{\pi} \int_d^\infty r^2 j^l(kr) V_l j^l(k^\prime r) dr,
\end{equation}
where the superscript $l$ refers to the angular momentum state. Figures \ref{1s0} and \ref{3s1a} compare the results of the Fourier transformation with $V_{\rm low-k}$. 
Of course, since the KK potential fits only the scattering length and effective range, it should not be expected to agree precisely with $V_{\rm low-k}$. Nevertheless, the agreement between the two appears generally good for the $^1S_0$ state but slightly worse for the $^3S_1$ state. It is not surprising that a local approximation is less good for the $^3S_1$ state than for the $^1S_0$ state, because a good fraction, \mbox{$\sim 1/3$}, of the $^3S_1$ attraction comes from the second order tensor interaction. The contributions peak quite sharply around intermediate states with momenta \mbox{$\sim 2$ ${\rm fm}^{-1}$} \cite{Brown}, so a local approximation is quite good, but to be completely local the peak would have to be a $\delta$-function.

\begin{figure}[htbp]
\centerline{\includegraphics[width=6.5cm, height=5.5cm]{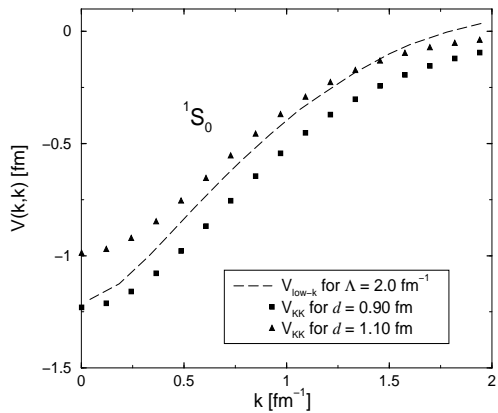}}
\caption{Variations of the $^1S_0$ partial wave matrix elements with the separation distance.}
\label{var}
\end{figure}

\begin{figure}[htbp]
\centerline{\includegraphics[width=6.5cm, height=5.5cm]{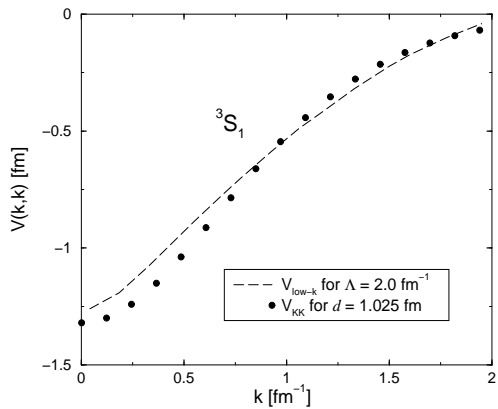}}
\caption{The $^3S_1$ diagonal matrix elements of $V_{\rm low-k}$ and the Kallio-Kolltveit potential for a configuration space cutoff of \mbox{1.025 fm}.}
\label{3s1b}
\end{figure}

We can gain some insight into the relationship between the MEEFT procedure and the separation method by examining the effects of varying the separation distance. Figure \ref{var} shows how the $k$-space KK potential compares to $V_{\rm low-k}$ for separation distances of \mbox{0.9 fm} and \mbox{1.1 fm} in the $^1S_0$ channel. 
It appears that for the $^1S_0$ state, a separation distance of \mbox{$\sim 1.0$ fm} produces the closest agreement with $V_{\rm low-k}$. Figure \ref{3s1b} shows the effect of raising the separation distance of the $^3S_1$ state to \mbox{$1.025$ fm}. The agreement with $V_{\rm low-k}$ is notably better. We suggest that this is explained by the common scale in $V_{\rm low-k}$.

\begin{figure}[htbp]
\centerline{\includegraphics[width=6.5cm, height=5.5cm]{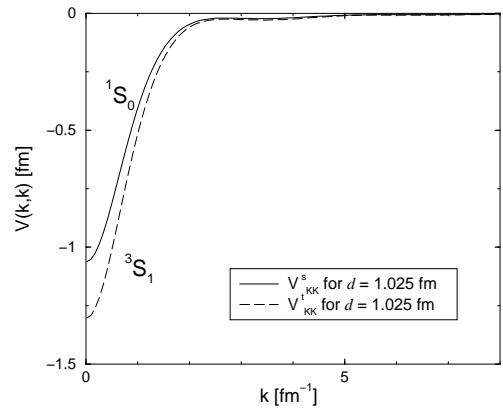}}
\caption{Diagonal matrix elements of the Kallio-Kolltveit potential, including momenta above the $V_{\rm low-k}$ cutoff of \mbox{2.0 $\rm fm^{-1}$}.}
\label{vhighk}
\end{figure}

The other question in comparison with $V_{\rm low-k}$ is as to the momentum components above $\Lambda=2.0$ $\rm fm^{-1}$, which are taken to be zero (modulo some artifacts from the cutoff) in the effective field theory because they have not been measured experimentally.  We show these momentum components for the KK potential in Figure \ref{vhighk}. From Figure \ref{vhighk} we see that the diagonal matrix elements are very small for $k > \Lambda \sim 2$ $\rm fm^{-1}$. Importantly, the sharp cutoff on the potential does not introduce appreciable artifacts.

We thus note that the $S$-wave treatment of Kuo and Brown \cite{Kuo}, modulo the small adjustment we made here to have equal cutoffs in singlet and triplet channels, using the MS separation method was equivalent to the MEEFT which results in $V_{\rm low-k}$. Since model dependence in terms of high-momentum Fourier components above those accessed in the nucleon-nucleon scattering experiments will occur predominantly in the $S$-wave channels, this gives an answer to why the Kuo-Brown interactions have endured for 38 years; namely, to a large extent these model-dependent momenta were not present in the KB interaction.

These comparisons seem to suggest first that the MEEFT and separation method predict the same separation of scales in the nuclear interaction. The fact that close agreement between $V_{\rm low-k}$ and the KK potential is reached at a separation distance of \mbox{$1.0$ fm} for both angular momentum states suggests that integrating out momenta beyond \mbox{$2.0$ $\rm fm^{-1}$} via the RG corresponds roughly to removing the short distance details below \mbox{$1.0$ fm}.

We are not suggesting a replacement for $V_{\rm low-k}$, which has been astonishingly successful in nuclear structure calculations. But we do show that with $S$-wave potentials with the usual schematic hard core of conventional \mbox{$0.4$ fm} radius, which fits the scattering length and effective range, we can get a good approximation, local in $r$, to $V_{\rm low-k}$ by choosing the separation distance correctly. It should be noted that the parameters in the KK potentials were chosen in order to get the scattering lengths and effective range correct. It thus appears that in addition to this some schematization of the short range repulsion is needed. With these minimal requirements one then has a good tool for nuclear structure physics. We suggest that these local potentials may be useful in schematic calculations where nuclear interactions have to be taken into account.

Our discussion here concerns only the G-matrix of Kuo and Brown \cite{Kuo}, where we show the $S$-wave interactions to be essentially those of $V_{\rm low-k}$. The important remaining question which occupied research workers for many years was the validity of the polarization bubble that they used. We plan to show in a future publication \cite{Bog4} using the Babu-Brown formalism \cite{Babu} which sums all planar particle-hole diagrams, that higher-order rescattering corrections reduce the strength of the bubble somewhat, especially at higher densities, but leave most of it. The higher-order corrections are only appreciable in the spin- and isospin-independent channels, affecting especially the compression modulus.

% Surround figure environment with turnpage environment for landscape
% figure
% \begin{turnpage}
% \begin{figure}
% \includegraphics{}%
% \caption{\label{}}
% \end{figure}
% \end{turnpage}
% tables should appear as floats within the text
%
% Here is an example of the general form of a table:
% Fill in the caption in the braces of the \caption{} command. Put the label
% that you will use with \ref{} command in the braces of the \label{} command.
% Insert the column specifiers (l, r, c, d, etc.) in the empty braces of the
% \begin{tabular}{} command.
% The ruledtabular enviroment adds doubled rules to table and sets a
% reasonable default table settings.
% Use the table* environment to get a full-width table in two-column
% Add \usepackage{longtable} and the longtable (or longtable*}
% environment for nicely formatted long tables. Or use the the [H]
% placement option to break a long table (with less control than 
% in longtable).
% \begin{table}%[H] add [H] placement to break table across pages
% \caption{\label{}}
% \begin{ruledtabular}
% \begin{tabular}{}
% Lines of table here ending with \\
% \end{tabular}
% \end{ruledtabular}
% \end{table}

% Surround table environment with turnpage environment for landscape
% table
% \begin{turnpage}
% \begin{table}
% \caption{\label{}}
% \begin{ruledtabular}
% \begin{tabular}{}
% \end{tabular}
% \end{ruledtabular}
% \end{table}
% \end{turnpage}
% Specify following sections are appendices. Use \appendix* if there
% only one appendix.
%\appendix
%\section{}

% If you have acknowledgments, this puts in the proper section head.
\begin{acknowledgments}
We thank T.T.S. Kuo and A. Schwenk for helpful discussions.
\end{acknowledgments}

% Create the reference section using BibTeX:
%\bibliography{basename of .bib file}

\end{document}